\documentclass[preprint,preprintnumbers,amsmath,amssymb]{revtex4}
\usepackage{amsfonts}    
\usepackage{amssymb}
\usepackage{latexsym}
\usepackage{eepic}
\usepackage{graphicx}
\usepackage[usenames,dvipsnames]{color}

\usepackage[active]{srcltx}

\newcommand{\al}{\alpha}
\newcommand{\pa}{\partial}
\newcommand{\ep}{\epsilon}

\newcommand{\la}{\lambda}
\newcommand{\ta}{\tau}

\newcommand{\de}{\delta}

\newcommand{\De}{\Delta}

\newcommand{\rar}{\rightarrow}

\begin{document}

\title{The Heun operator as a Hamiltonian}

\author{\large Alexander~V.~Turbiner}
\email{turbiner@nucleares.unam.mx}

\bigskip
\affiliation{Instituto de Ciencias Nucleares, Universidad Nacional
Aut\'onoma de M\'exico, Apartado Postal 70-543, 04510 M\'exico,
D.F., Mexico}
%

\bigskip

\begin{abstract}
It is shown that the celebrated Heun operator
$H_e=-(a_0 x^3 + a_1 x^2 + a_2 x) \frac{d^2}{dx^2} + (b_0 x^2 + b_1 x + b_2)\frac{d}{dx}
+ c_0 x$ is the Hamiltonian of the $sl(2,R)$-quantum Euler-Arnold top of spin $\nu$ in a constant magnetic field.
For $a_0 \neq 0$ it is canonically-equivalent to $BC_1(A_1)-$ Calogero-Moser-Sutherland quantum models, if $a_0=0$, ten known one-dimensional quasi-exactly-solvable problems are reproduced, and if, in addition, $b_0=c_0=0$, then four well-known one-dimensional quantal exactly-solvable problems are reproduced. If spin $\nu$  of the top takes (half)-integer value the Hamiltonian possesses a finite-dimensional invariant subspace and a number of polynomial eigenfunctions occurs.
Discrete systems on uniform and exponential lattices are introduced which are canonically-equivalent to one described by the Heun operator.
\end{abstract}


\hskip 2cm


\maketitle

\section{The Heun operator}

Needless to say that the Heun operator $P_3 \pa_x^2 + P_2 \pa_x + P_1$ plays exceptionally important role in different physical sciences and mathematics, see e.g. the recent papers \cite{Luc} and \cite{Eremenko}, and some references therein, respectively. It is characterized by four regular singular points and has the form
\begin{equation}
\label{heun}
  H_e\ =\ -(a_0 x^3 + a_1 x^2 + a_2 x) \pa_x^2 + (b_0 x^2 + b_1 x + b_2) \pa_x + c_0 x\ ,\quad \pa_x \equiv \frac{d}{dx}\ ,
\end{equation}
\bigskip
it depends on 7 free parameters $(a_{0,1,2}, b_{0,1,2}, c_0)$:
we do not fix normalization (overall factor) but choose the reference point for $x$
in such a way that the coefficient $P_3$ in front of the second derivative vanishes at $x=0$;
for a sake of future convenience, we do {\it not} factorize the coefficient $(a_0 x^2 + a_1 x + a_2) x$ further to a product of monomials. The operator (\ref{heun}) is defined up to additive constant $c_1$ - it is the reference point for spectral parameter and coincides with the accessory parameter in the Heun equation.

In the basis of monomials the operator $H_e$ has a form of tridiagonal matrix. Note that the celebrated Heun equation \cite{Heun} (see for the general discussion e.g. \cite{Ronveaux}) and also \cite{Maier}, \cite{Luc}, \cite{Eremenko}
\[
     \frac{1}{P_3}\,(H_e - c_1)\, u(x) \ =\ 0\ ,
\]
occurs, where $c_1$ is the accessory parameter.

\section{The Euler-Arnold quantum top}

The Hamiltonian of the $sl(2,R)$-quantum Euler-Arnold top of spin $\nu$ in a constant magnetic field ${\bf B}$ is given by
\begin{equation}
\label{top}
  H\ =\ t^{+0} J_+ J_0 + t^{+-} J_+ J_- + t^{00} J_0 J_0 + t^{0-} J_0 J_- + B^+ J_+ + B^0 J_0 + B^- J_-\ ,
\end{equation}
where $t^{+0}, t^{+-}, t^{00}, t^{0-}$ and $B^+, B^0, B^-$ are constants. Here $J$'s span the $sl(2,R)$-Lie algebra: they obey the $sl(2,R)$ commutation relations: making linear transformations the general $sl(2,R)$ quantum top in a constant magnetic field can be reduced to the form (\ref{top}). It is evident that this top is integrable: the second Casimir operator $C_2$ is the integral, it always commutes with $H$.
The top (\ref{top}) can be constrained by imposing a condition that $C_2$ is a constant,
\begin{equation}
\label{c2}
C_2 \equiv \frac{1}{2} \{J_+\, ,\,J_-\} -  J_0 J_0\ =\ - \nu (\nu + 1)\ ,
\end{equation}
where $\{J_+\, ,\, J_-\}\ \equiv\ J_+ J_- + J_- J_+$ denotes the anti-commutator.
Here $t^{\al,\beta}$\ with\, $\al,\beta=\pm,0$\ defines the tensor of inertia, ${\bf B}=(B^+, B^0, B^-)$ defines a magnetic field. Thus, the Hamiltonian (\ref{top}) with constraint (\ref{c2}) depends on 7 free parameters as well as the operator (\ref{heun}).

\section{\large \bf $sl(2,R)$-Lie algebra}

The $sl(2,R)$-algebra can be naturally spanned by differential operators, by difference (shift) operators and by difference (dilation) operators.

\subsection{$sl(2,R)$-Lie algebra and differential operators}

In first order differential operators the $sl(2,R)$-Lie algebra is realized as follows,
\begin{equation}
\label{sl2}
    J_-\ =\ {\pa_x}\quad ,\quad J_0\ =\ x {\pa_x} - \nu\ ,\
    J_+\ =\ x (x {\pa_x} - 2\nu)\ ,
\end{equation}
where the parameter $\nu$ is spin. It is easy to see that in representation (\ref{sl2}) the Casimir operator $C_2$ is the constant, see (\ref{c2}), thus, the constraint is fulfilled.

By substituting (\ref{sl2}) into (\ref{top}) we get the Heun operator (\ref{heun}), hence, $H=H_e$. In particular,
\begin{equation}
\label{connect}
    -a_0=t^{+0}\ ,\ b_0=t^{+0} (1 - 3\nu)+B^+\ ,\ c_0=2\nu(\nu t^{+0}-B^+)\ .
\end{equation}
These parameters satisfy a remarkable condition
\begin{equation}
\label{cond}
 -2\nu(2\nu-1) a_0 + 2\nu b_0 + c_0\ =\ 0\ ,
\end{equation}
which implies that for any $a_0, b_0, c_0$ one can find $\nu$ such that this condition is fulfilled. This $\nu$ has a meaning of the $sl(2,R)$ spin of representation. One can reparameterize (\ref{heun}) by replacing
\[
    c_0\ =\ 2\nu(2\nu-1) a_0 - 2\nu b_0 \ ,
\]
to have the condition (\ref{cond}) fulfilled automatically. It is evident that the Heun operator acts in infinite-dimensional space of monomials $\{\ldots x^{-n+\nu}, \ldots x^{-1+\nu}, x^{\nu} \}$.

\subsection{$sl(2,R)$-Lie algebra and difference (shift) operators}

Let us introduce the shift operator,
\[
      T_{\de} f(x)\ =\ f(x + \de)\quad ,\quad T_{\de}=e^{\de \pa_x}\ ,
\]
and construct a canonical pair of shift operators (see e.g. \cite{Turbiner:1995})
\begin{equation}
\label{delta}
D_{\de}\ =\ \frac{T_{\de} - 1}{\de}\ ,\ x_{\de}\ =\ x T_{-\de} = x(1-\de{ D}_{-\de}) \ ,
\end{equation}
where the operator $D_{\de}$,
\[
        D_{\de} f(x) = \frac{f(x + \de) - f(x)}{\de}\ ,
\]
is the so-called Norlund derivative. It is easy to check that $[D_{\de}, x_{\de}]=1$, hence, forming the canonical pair.

In first order difference (shift) operators the $sl(2,R)$-Lie algebra is realized as follows,
\begin{equation}
\label{sl2-s}
    J_-\ =\ D_{\de} \quad ,\quad J_0\ =\ x_{\de} D_{\de} - \nu\ ,\
    J_+\ =\ x_{\de} (x_{\de} D_{\de} - 2\nu)\ ,
\end{equation}
where the parameter $\nu$ is spin, while $\de$ is arbitrary parameter. It is easy to see that in representation (\ref{sl2-s}) the Casimir operator $C_2$ is the constant, see (\ref{c2}), it does not depend on $\de$, thus, the constraint is fulfilled.
By substituting (\ref{sl2-s}) into (\ref{top}) we get the finite-difference (shift) Heun operator
\begin{equation}
\label{heun-s}
  H^{(s)}_e\ =\ -(a_0 x_{\de}^3 + a_1 x_{\de}^2 + a_2 x_{\de}) D_{\de}^2 + (b_0 x_{\de}^2 + b_1 x_{\de} + b_2) D_{\de} + c_0 x_{\de}\ ,
\end{equation}
cf. (\ref{heun}). This operator acts on uniform lattice space with spacing $\de$
\[
 \{ \ldots,\ x-2\de\ ,\ x -\de\ ,\ x\ ,\ x+\de\ ,\ x+2\de\ ,\ \ldots  \}
\]
marked by $x \in {\bf R}$ - a position of a central (or reference) point of the lattice. Explicitly, it is five-point lattice operator,
\[
   H^{(s)}_e\,f(x)\ =\ A_{-3}(x) f(x-3\de) + A_{-2}(x) f(x-2\de) +  A_{-1}(x) f(x-\de) +
   A_{0}(x) f(x) + A_{1}(x) f(x+\de) \ ,
\]
where $A$'s are polynomials. In the limit $\de \rar 0$, the operator $H^{(s)}_e$ becomes the Heun operator $H_e$ (\ref{heun}).

\subsection{$sl(2,R)$-Lie algebra and difference (dilation) operators}

Let us introduce the dilation operator,
\[
      T_{q}\, f(x)\ =\ f(q x)\quad ,\quad T_{q}=q^{A}\ ,\ A \equiv x\, \pa_x\ ,
\]
and construct a canonical pair of dilation operators \cite{CT}
\begin{equation}
\label{q}
D_{q}\ =\ x^{-1}\frac{T_{q} - 1}{q-1}\ ,\ x_{q}\ =\ \frac{A (q-1)}{T_{q} - 1}\, x \ ,
\end{equation}
where $x_{q}\,D_{q}=A$ and $D_{q}\, x_{q}=\pa_x\, x=A+1$. The operator $D_q$ is the so-called Jackson symbol (or the Jackson derivative).
Both operators $x_{q}\ ,\ D_{q}$ are pseudodifferential operators with action on monomials as follows,
\[
        D_{q} x^n = \{ n \}_q\, x^{n-1}\ ,\ x_q x^n\ =\ \frac{n+1}{\{ n+1\}_q}\,x^{n+1}\ ,
\]
where $\{ n \}_q=\frac{1 - q^n}{1 - q}$ is the so called $q$-number.

In first order difference (dilation) operators the $sl(2,R)$-Lie algebra is realized in the following way,
\begin{equation}
\label{sl2-d}
    J_-\ =\ D_{q}\quad ,\quad J_0\ =\ x {\pa_x} - \nu\ ,\
    J_+\ =\ x_q (x {\pa_x} - 2\nu)\ ,
\end{equation}
cf.(\ref{sl2}), where the parameter $\nu$ is spin. It is easy to see that in representation (\ref{sl2-d}) the Casimir operator $C_2$ is the constant, see (\ref{c2}), it does not depend on $q$, thus, the constraint is fulfilled.
By substituting (\ref{sl2-d}) into (\ref{top}) we get the finite-difference (dilation) Heun operator
\begin{equation}
\label{heun-d}
  H^{(d)}_e\ =\ -(a_0 x_{q}^3 + a_1 x_{q}^2 + a_2 x_{q}) D_{q}^2 + (b_0 x_{q}^2 + b_1 x_{q} + b_2) D_{q} + c_0 x_{q}\ ,
\end{equation}
cf. (\ref{heun}), (\ref{heun-s}). In fact, it is the differential-difference operator
\[
   H^{(d)}_e\ =\ -(a_0 x_{q}^2 x \pa_x + a_1 x_{q} x \pa_x + a_2 x \pa_x ) D_{q} + (b_0 x_{q} x \pa_x + b_1 x \pa_x + b_2 D_{q}) + c_0 x_{q}\ ,
\]
acting on exponential lattice $\{ x q^n \}\ ,\ n=0,1,2,\ldots$ with reference point marked by $x \in {\bf R}$.

\section{Properties}

\begin{itemize}
  \item Operators $H_e, H^{(s)}_e, H^{(d)}_e$ are canonically-equivalent, see (\ref{delta}), (\ref{q});

  \item  if $2 \nu=n$ is integer, the Heun operator $H_e (H^{(s)}_e, H^{(d)}_e)$ (\ref{heun}) ((\ref{heun-s}), (\ref{heun-d})) has $(n+1)$-dimensional invariant subspace in polynomials ${\cal P}_n$, hence,
      $[H_e, \pa_x^{n+1}]: {\cal P}_{n} \rar \{ 0 \}$ and $\pa_x^{n+1}= (J_-)^{n+1}$ is particular integral \cite{Turbiner:2013}, also $[H^{(s,d)}_e, (J_-)^{n+1}]: {\cal P}_{n} \rar \{ 0 \}$, see (\ref{sl2-s}), (\ref{sl2-d});

  \item ``Polynomial" isospectrality. Let us assume the Heun operator $H_e (x)$ has a formal polynomial eigenfunction,
\[
        \phi (x) \ =\ \sum_{k=0}^{n} \al_k x^k \ ,
\]
with eigenvalue $\ep$, then the discrete (shift) Heun operator $H^{(s)}_e (x)$ has a formal polynomial eigenfunction
\[
        \phi_{\de} (x) \ =\ \sum_{k=0}^{n} \al_k x^{(k)} \ ,
\]
where $x^{(k+1)} \equiv x(x-\de)\ldots (x - k \de)$ is the Pochhammer symbol or {\it quasi-monomial},
with eigenvalue $\ep$ and the discrete (dilation) Heun operator $H^{(d)}_e (x)$ has a formal polynomial eigenfunction
\[
        \phi_q (x) \ =\ \sum_{k=0}^{n} \al_k \frac{k!}{\{k\}_q!} x^k \ ,
\]
where $k!=1 \cdot 2 \cdot \ldots k$ is factorial and $\{ k \}_q! = \{ 1 \}_q \cdot \{ 2 \}_q \cdot \ldots \{ k \}_q$ is the so-called $q$-factorial,
with eigenvalue $\ep$;

  \item  making canonical (gauge rotation) transformation
\begin{equation}
\label{can1}
    \pa_x \rar \pa_x + {\cal A}\ \equiv\ e^{-\phi}\pa_x e^{\phi} \ ,\ x \rar x\ ,
\end{equation}
where ${\cal A}=\phi'(x)$, the Heun operator (\ref{heun}) can be transformed to the Schr\"odinger operator with rational potential
\begin{equation}
\label{sch-g}
    H_e(x) = -\De_g(x) + V(x)\ ,\ V(x)\ =\ B x + \frac{Q_2(x)}{P_3(x)}\ ,
\end{equation}
where $\De_g(x)$ is one-dimensional Laplace-Beltrami operator with contravariant metric $g^{11} = P_3$,
$B = \frac{3}{16}a_0+23 b_0+c_0+\frac{b_0^2}{4a_0}$ is the parameter, $Q_2$ is a 2nd degree polynomial.
Then making a change of variables - another canonical transformation,
\begin{equation}
\label{can2}
    \pa_x \rar \frac{1}{x'_{\ta}}\pa_{\ta} \ ,\ x \rar \ta(x)\ ,
\end{equation}
where $(x_{\ta}')^2 = a_0 x^3 + a_1 x^2 + a_2 x$, we arrive at the Schr\"odinger operator in the Cartesian coordinate $\ta$,
\begin{equation}
\label{sch-f}
    H_e(\ta) = -\pa_{\ta}^2 + V(x(\ta))\ ,\ V(x(\ta))\ =\ B x(\ta) + \frac{Q_2(x(\ta))}{P_3(x(\ta))}\ ,
\end{equation}

  \item ``Canonical" covariance. Take $(x-\al)^{\mu}$ as the gauge factor, where $\al$ is root of the cubic equation $P_3(\al)=0$. For any parameters $\{a,b,c \}$ one can indicate $\mu$ such that the gauge-rotated Heun operator remains the Heun operator,
\[
      (x-\al)^{-\mu}\ H_e(x;\{a,b,c \})\ (x-\al)^{\mu}\ =\ H_e(x;\{ a, {\tilde b}, {\tilde c} \})\ .
\]
For example, for zero root, $\al=0$, the exponent $\mu=1+\frac{b_2}{a_2}$. For the Lame operator $\mu=\frac{1}{2}$ for any root $\al$

  \item  if $a_0 \neq 0$,

  {\bf (i)}\ $H_e$ with additive constant $c_1$ is factorizable in the $sl(2,R)$-algebra
\[
   H_e + c_1 = T_a T_b\ ,\
\]
  where $T_{a,b} = \al_{a,b} J_+ + \beta_{a,b} J_0 + \gamma_{a,b} J_- + D_{a,b}$ with $\al_b=0$ and $c_1=D_a D_b$; \\
  {\bf (ii)}\ $H_e$ is canonically-equivalent to the $BC_1$ elliptic Inozemtsev Hamiltonian \cite{Inozemtsev:1989} (see \cite{Takemura:2002}), to the $BC_1 (A_1)$ elliptic Calogero-Moser-Sutherland Hamiltonian (see e.g. \cite{Olshanetsky:1983}) written in Cartesian coordinate $\ta$; if in (\ref{heun}), $P_2 = -P_3'/2$, $H_e$ coincides with the Lame operator.
\end{itemize}
{\bf Example: \it $BC_1$ elliptic Calogero-Moser-Sutherland model}.

Take the Heun operator
\[
   h_{BC_1}(x)\ =\
\]
\begin{equation}
\label{hBC1}
    \ 4({x}^3 - 3 \, \la {x}^2\ +\ 3 \de {x}) \pa^2_{{x}} \ +\
    6 (1+2\mu) ({x}^2-2 \la {x}+\de) \pa_{{x}}\ -\ 2n(2n+1+6\mu) ({x} - \la)\ ,
\end{equation}
where $\la, \de, \mu, n$ are parameters; here $a_0=-4$, $b_0=6 (1+2\mu)$, $c_0=-\ 2n(2n+1+6\mu)$,\, c.f. (\ref{heun}) and $n=2\nu$, c.f. (\ref{cond}).
In space of monomials $x^k,\ k=0,1,2,\ldots$ the operator $h_{BC_1}(x)$ has the form of tri-diagonal, Jacobi matrix. In terms of ${sl}(2,{R})$-generators (\ref{sl2}),
it reads
\[
 {h}_{BC_1}\ =\ 4\ { J}^+(n)\, { J}^0(n)\ -\ 12\la {J}^0(n)\, {J}^0(n)\ +
 \ 12\de {\ J}^0(n)\, { J}^-
\]
\begin{equation}
\label{hBC1-sl2}
 +\ 2\,(4n + 1 + 6\mu)\,{J}^+(n)\ -\ 12\,\la (n+2\mu) J^0(n)\ +\ 6\,\de\,(n+1+ 2\mu)\,{J}^-
 \ +\ \la n(n+2) \ .
\end{equation}
Making the gauge rotation of (\ref{hBC1}),
\[
  {\mathcal H}_{\rm BC_1}\ =\ -\frac{1}{2}\, (\Psi_{0})\,{h}_{BC_1}\,\Psi_{0}^{-1}\ ,\
  \Psi_{0}\ =\ [\ {x}^3 - 3 \, \la {x}^2\ +\ 3 \de {x}\ ]^{\frac{\mu}{2}}\ ,
\]
and changing $ x \rar \ta$,
\[
     (x_{\ta}')^2 = 4({x}^3 - 3 \, \la {x}^2\ +\ 3 \de {x})\ ,
\]
we arrive at $BC_1$ elliptic Calogero-Moser-Sutherland Hamiltonian \cite{Turbiner:2015},
\begin{equation}
\label{HBC1}
    {\mathcal H}_B\ =\ - \frac{1}{2}\, \pa^2_{{\ta}} \ +\ 2\mu(\mu-1)\ \wp (2\ta)\ +\ (2n+1+2\mu)(n+2\mu)\ \wp (\ta)
    \ ,
\end{equation}
where $\wp (\ta)\equiv \wp(\ta|g_2,g_3)$ is the Weierstrass function with its invariants parametrized as follows,
\[
     g_2 = 12 (\la^2 - \de)\ ,\qquad  g_3=4 \la (2\la^2-3\de)\ .
\]
Here $\la, \de$ are parameters and $e=-\la$ is the root of the $\wp-$Weierstrass function, $\wp'(-\la)=0$. In such a parametrization the variable $x$ has the form
\[
    x\ =\ \wp(\ta|g_2,g_3)+\la \ .
\]

\begin{itemize}
  \item  if $a_0=0$, but $b_0 > 0$, all known ten families of QES Hamiltonians
  written in Cartesian coordinate $\ta$ are reproduced \cite{Turbiner:1988a,Turbiner:1988b,Turbiner:1992}.

  \item  if $a_0=0$, and $b_0=c_0=0$, the Riemann (hypergeometrical) operator occurs,
  thus, all known quantal exactly-solvable problems (the Harmonic oscillator, the Morse and the P\"oschl-Teller potentials, the Coulomb problem) are reproduced when written in the Cartesian coordinate $\tau$
  \, .

\end{itemize}

\section{Classical limit}

The classical limit appears when in the top Hamiltonian (\ref{top}) and constraint (\ref{c2}) the generators $J$'s span the $sl(2,R)$ Poisson algebra, thus, the Lie bracket is replaced by the Poisson bracket. In this case we have the classical $sl(2,R)$ top in a constant magnetic field.
Hence, the Hamiltonian is the polynomial in coordinate and momentum of the third and second  degrees, respectively. Following the Igor Krichever suggestion we call this procedure {\it de-quantization}. Classical version of the Hamiltonian (\ref{sch-g}) based on de-quantization, $\pa_x \rar i p, x \rar q$, is characterized by rational potential,
\begin{equation}
\label{class-g}
    {\cal H}_e(q,p)\ =\ P_3(q)\, p^2\ +\ B\, x\ +\ \frac{Q_2(q)}{P_3(q)}\ ,
\end{equation}
cf. (\ref{sch-g}), where $p$ is the classical momentum. Classical trajectories (in phase space), ${\cal H}_e=E$, are algebraic curves.

\section{Conclusions}

In this Letter we show that the Heun operator is the Hamiltonian of the $sl(2,R)$-(non-compact) Lie algebra, quantum Euler-Arnold top in a constant magnetic field. It reveals the links on the level of quantum canonical equivalence between three different quantum systems: tops, (generalized) Calogero-Moser-Sutherland systems and discrete systems, all in one dimension. It implies that in the Fock space (Universal Enveloping Heisenberg Algebra $U_{H_3}$ with vacuum attached) formalism all three systems are described by a single polynomial in $H_3$ generators. There are clear indications that a similar links exist in multidimensional case involving $A_n, BC_n, D_n$ quantum Calogero-Moser-Sutherland models and $sl(n+1,R)$-algebra quantum Euler-Arnold top, $G_2$ quantum Calogero-Moser-Sutherland model and $g^{(2)}$-polynomial algebra, quantum top (see \cite{Sokolov:2015} for $A_2/G_2$ case), and also other Calogero-Moser-Sutherland models and polynomial algebra, quantum tops. These links will be described elsewhere.

\section*{ Acknowledgements}

The research is supported in part by DGAPA grant IN108815 (Mexico). The author thanks
A Abanov (Stony Brook), A Eremenko (Purdue), N Nekrasov (Stony Brook) and, especially, S P Novikov (Maryland) for useful conversations. I am also grateful to S P Novikov for the encouragement to write the extended Conclusions.

The author gratefully acknowledges support from the Simons Center for Geometry and Physics, Stony Brook University at which some of the research for this paper was performed.

\end{document}